\documentclass[aps,prd,groupedaddress,nofootinbib]{revtex4}

\def\ga{\mathrel{\raise.3ex\hbox{$>$\kern-.75em\lower1ex\hbox{$\sim$}}}}
\def\la{\mathrel{\raise.3ex\hbox{$<$\kern-.75em\lower1ex\hbox{$\sim$}}}}

\def\lsim{\mathrel{\rlap{\lower4pt\hbox{\hskip1pt$\sim$}}
    \raise1pt\hbox{$<$}}}                
\def\gsim{\mathrel{\rlap{\lower4pt\hbox{\hskip1pt$\sim$}}
    \raise1pt\hbox{$>$}}}                

\usepackage{amsmath}
\usepackage{amssymb}
\usepackage{bbm}
\usepackage{epsfig}
\usepackage{slashed}

\begin{document}
\begin{flushright}
SHEP-09-12\\
DFTT-60-2009
\end{flushright}
\title{Very Light Higgs Bosons in Extended Models at the LHC}

\author{Alexander Belyaev$^{1}$~\footnote{a.belyaev@soton.ac.uk} Renato Guedes$^{1}$~\footnote{r.b.guedes@soton.ac.uk},
Stefano Moretti$^{1,2}$~\footnote{stefano@phys.soton.ac.uk},
and Rui Santos$^{1}$~\footnote{rsantos@cii.fc.ul.pt}}
\affiliation{$^1$NExT Institute and School of Physics and Astronomy,
University of Southampton, Highfield, Southampton SO17 1BJ, UK.}

\affiliation{$^2$Dipartimento di Fisica Teorica, Universit\`a di Torino,
Via Pietro Giuria 1, 10125 Torino, Italy}

\date{\today}

\begin{abstract}
\noindent
The Large Electron-Positron (LEP) collider experiments have constrained the mass of the Standard Model (SM) Higgs boson to be above 114.4 $GeV$. This bound applies to all extensions of the SM where the coupling of a Higgs boson to the $Z$ boson and also the Higgs decay profile do not differ much from the SM one. However, in scenarios with extended Higgs sectors, this coupling can be made very small by a suitable choice of the parameters of the model. In such cases, the lightest CP-even Higgs boson mass can in turn be made very small. Such a very light Higgs state, with a mass of the order of the $Z$ boson one or even smaller, could have escaped detection at LEP. In this work we perform a detailed parton level study on the feasibility of the detection of such a very light Higgs particle at the Large Hadron Collider (LHC) in the production process $pp \to h j \to \tau^+ \tau^- j$, where $j$ is a resolved jet. We conclude that there are several models where such a Higgs state could be detected at the LHC with early data.
\end{abstract}
%
\maketitle

\section{Introduction}
\label{sec:intr}

The Higgs mechanism gives rise to clear and detectable signatures that can be probed at hadron colliders. The Tevatron is now leading a race~\cite{Collaboration:2009je}, that the LHC will soon join, to find this spinless particle. The most relevant searches for a SM Higgs boson by the LEP experiments~\cite{Schael:2006cr} were based on the associated production mechanism, via $e^+ e^- \to Z h( \to b \bar{b})$ and $e^+ e^- \to Z h (\to \tau^+ \tau^-)$. A limit of $m_h > 114.4$ $GeV$ was obtained combining all LEP analyses based on such searches. As the coupling of a Higgs boson to gauge bosons is fixed in the SM, it is clear that a lighter Higgs can only exist in a model where its couplings to gauge bosons (and possibly those to fermions) are reduced relative to the SM. Moreover, any complementary process, such as $e^+e^-\to Ah$, where $A$ is a pseudo-scalar Higgs boson and $h$ is the lightest Higgs scalar boson (i.e., with the same quantum number as the SM state) in the model, which appears, e.g., in a pure 2-Higgs Doublet Model (2HDM) or in the Minimal Supersymmetric Standard Model (MSSM), has to be kinematically forbidden. This is a consequence of the obvious sum rule $g_{hZZ}^2 + g_{hAZ}^2$ valid in a pure 2HDM and in the MSSM. There were also searches at LEP based on the Yukawa processes $e^+ e^- \to b \bar{b} h(\to \tau^+ \tau^-)$ in the Higgs mass range $m_h =4-12$ GeV in~\cite{Abbiendi:2001kp} and in the channels $b \bar{b} b \bar{b}$, $b \bar{b} \tau^+ \tau^-$ and $ \tau^+ \tau^- \tau^+ \tau^-$ for Higgs masses up to 50 GeV in~\cite{Abdallah:2004wy}. These studies are only relevant for large values of the Yukawa coupling constants and will be discussed in section~\ref{sec:bounds}.

Such a scenario, with a very light Higgs, can easily arise by adding Higgs scalar singlets and/or doublets to the Higgs sector of the SM. Starting from the first simple extension that allows for a lighter Higgs state - i.e., adding a neutral singlet - we will explore all possible scenarios to a maximum of a Democratic 3-Higgs Doublet Model (3HDM (D))~\cite{Grossman:1994jb} plus one neutral scalar singlet. We will see that from the very simplest extension to all models built thereafter, a very light Higgs state is allowed via a reduction of the couplings to gauge bosons. As expected, the more freedom is added by the new fields the more parameter space is available to accommodate a light Higgs boson. The possibility of the existence of a very light Higgs boson, with a mass below 80 GeV, in the context of models with two Higgs doublets, was discussed in~\cite{Kalinowski:1995dm}. More recently, a very light Higgs boson in the context of the MSSM, with a mass as low as about 60 GeV, was discussed in~\cite{Belyaev:2006rf}.

Assuming this light object would have eluded all LEP searches due to a small enough $g_{hVV}$  coupling (where $V=W,Z$), the production mechanisms that involve such $g_{hVV}$ couplings, like vector boson fusion or Higgs-strahlung, have consequently negligible cross sections. Therefore, we have to rely on the production modes where Higgs couplings to fermions are present. Under these conditions, the process with the largest cross section is gluon fusion. Hence, we have performed a detailed parton level study on the feasibility of the detection of a very light Higgs state (below $\sim 100 \, GeV$) at the LHC in the production process $pp \to h j \to \tau^+ \tau^- j$~\cite{Ellis:1987xu}, where $j$ represents a resolved jet (that indeed proceeds mainly via gluon fusion). We have recently presented a similar study for the SM Higgs boson~\cite{us0} (see also \cite{Mellado:2004tj}) for the LHC and a study for the Tevatron was performed in~\cite{Belyaev:2002zz}.

The reason to choose such a final state, involving $\tau$'s, amongst those accessible at hadron colliders, is clear. Had we chosen $h \to b \bar{b}$ instead, we would have been overwhelmed by QCD background while the final state with $h \to \gamma \gamma$ was not considered because when we vary the Higgs boson mass from 120 $GeV$ (where the Branching Ratio (BR) roughly peaks for a SM-like Higgs) down to 20 $GeV$, the corresponding BR drops by a factor of $\sim$ 10. Therefore, unless one is exploring a model where the Higgs decay to photons is enhanced, the $h \to \tau^+ \tau^-$ mode is the most appropriate channel for light Higgs states. Other production channels, like $pp \to h t \bar{t}$, will be explored in the future~\cite{us}. We will show that such a very light Higgs could be detected with this process in several extended models and that for particular scenarios an early detection is also possible. At the very least, an effort should be made to definitely exclude such a light particle and the LHC definitely has the means to do so.

The plan of the paper is as follows. The next section is devoted to describe the signal and background processes in the SM while the following one extrapolates our findings to a variety of beyond the SM scenarios with an enlarged Higgs sector. Sect. IV introduces the experimental and theoretical bounds enforced in our analysis. Final results are presented in Sect. V (for various scenarios separately) while Sect. VI draws our conclusions. Finally, one appendix will help us in classifying four types of 2HDM.

\section{Signal and backgrounds in the SM}
\label{sec:signal}

In a recent work~\cite{us0} we have performed a detailed parton level study on the feasibility of the detection of a Higgs boson in the gluon fusion process $pp (gg+gq) \to  h j\to  \tau^+ \tau^- j$  at the LHC. In this section we will extend the study for Higgs masses below 100 $GeV$ to probe extensions of the SM where a very light Higgs is still allowed. The results will be presented considering the case
where all Higgs couplings to fermions are the SM ones so that they can be used in extensions of the SM with the same final state. As a detailed discussion of the parton level analysis was already presented in~\cite{us0}, here we will just highlight the main points and refer the reader to reference~\cite{us0} for details. The signal, $pp \to gg (q) \to hg (q) \to \tau^+ \tau^- g (q)$, is a one loop process with partonic contributions from
$gg \to hg$, $gq \to hq$, which is approximately 20 \% of the total cross section, and $qq \to hg$, which was shown to be negligible~\cite{Abdullin:1998er} and was not taken into account in our study. We note that this is a parton level study: effects of initial and final state radiation as well as hadronisation were not taken into account.

The SM signal and all background processes were generated with CalcHEP~\cite{Pukhov:2004ca} and cross checked with MadGraph/MadEvent~\cite{Alwall:2007st}. The Higgs BRs to $\tau^+ \tau^-$ were evaluated with the HDECAY~\cite{Djouadi:1997yw} package (and modifications thereof). In the models with an extended scalar sector, the one loop amplitudes for the signal $pp \to gg (q) \to hg (q) \to \tau^+ \tau^- g (q)$ were generated and calculated with the packages FeynArts~\cite{feynarts} and FormCalc~\cite{formcalc}. The scalar integrals were evaluated with  LoopTools~\cite{looptools} and the CTEQ6L parton distribution functions~\cite{cteq} were used. The jet (leptons) energies were smeared according to the following Gaussian distribution
\begin{equation}
\frac{\Delta E}{E}=\frac{0.5 (0.15)}{\sqrt{E}} \, GeV,
\end{equation}
to take into account the respective detector energy resolution effects, where $0.5$ is the factor for jets while $0.15$ is the corresponding factor for leptons.

Each $\tau$ can either decay leptonically or hadronically. As a three jet final state is very hard to identify at a hadron collider, we will concentrate on the other two possibilities - two taus decaying leptonically ($ll$) or one tau decaying leptonically  and the other hadronically ($lj$). These are also the final states with robust trigger signatures - the events are selected by an isolated electron with $p_T^e > 22 \, GeV$  or an isolated muon with $p_T^\mu > 20 \, GeV$. In both analyses we have considered the main source of irreducible background: $pp \to Z/\gamma^* j \to l l j$  for $ll$ and $pp \to Z/\gamma^* j \to l j j$, where one jet originates from a tau, for the $l j$ case. In $pp \to Z/\gamma^* j \to l l j$ we include all possible combinations of $l=e,\mu$ and in $pp \to Z/\gamma^* j \to l j j$ only the intermediate state $\tau^+ \tau^- j$ is included - the $jjj$ signature, where a jet would fake a lepton with a given probability, is taken into account in the $jjj$ background.

The main source of reducible background for the $ll$ analysis comes from $pp\to W^+ W^- j$ while for the $lj$ case it is the process $pp\to W j j$ that dominates~\cite{us0}. The tau reconstruction efficiency was taken to be 0.3 and accordingly we have used a tau rejection factor against jets as a function of the jet $p_T$ using the values presented in the ATLAS study in~\cite{Aad:2009wy}. Finally, we have included the $pp \to t\bar{t}$ background taken at Next-to-Leading-Order (NLO). By vetoing the events if the tagging jet is consistent with a $b$-jet hypothesis for $|\eta| < 2.5$ we were able to discard most of the $t\bar{t}$ background. The $t\bar{t}$ background is larger for $ll$ as there are more possible combinations when the $W$ bosons decay leptonically.

\begin{widetext}
\begin{figure}[h]
\begin{center}
\includegraphics[width=7.5cm]{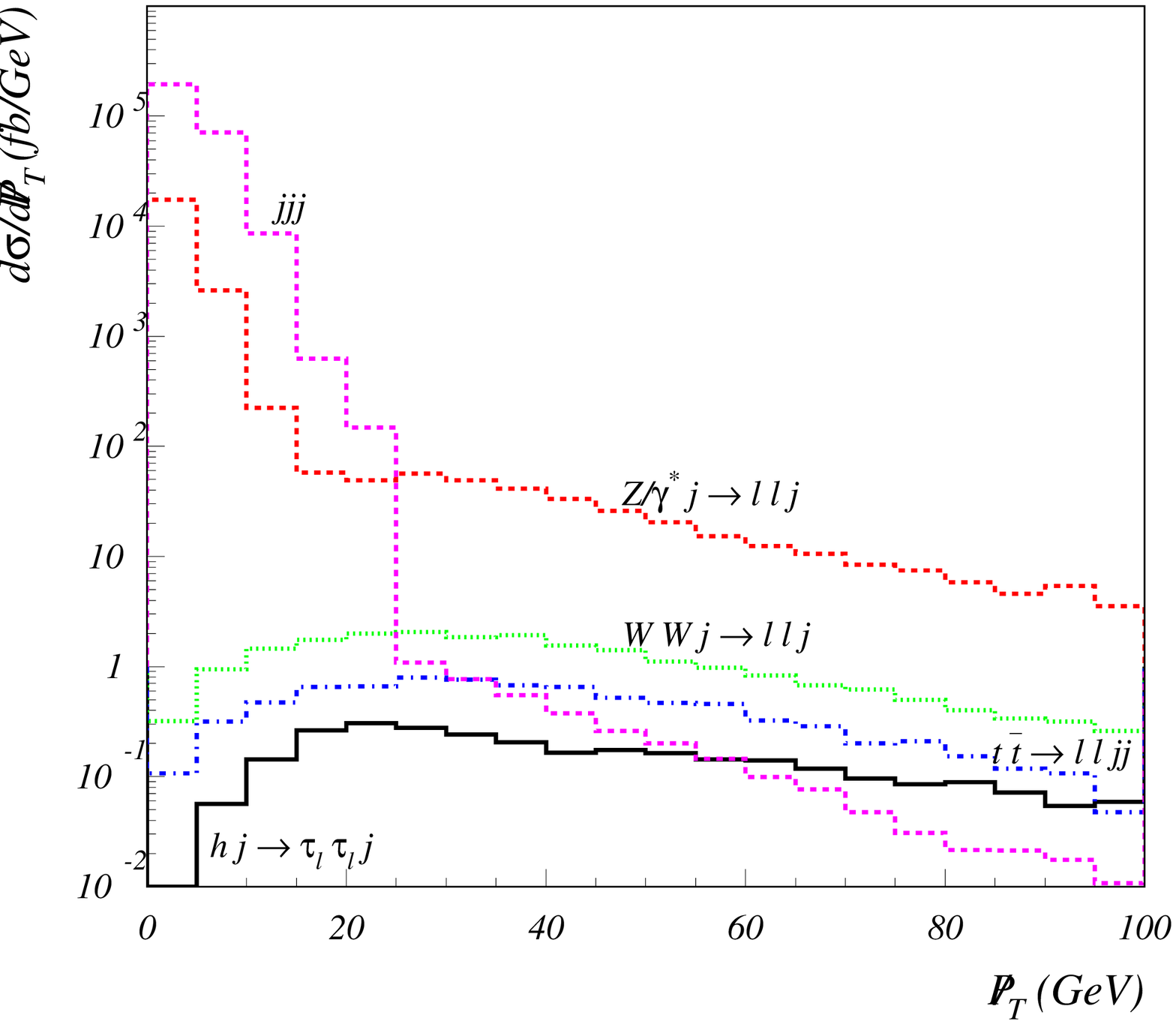}
\includegraphics[width=7.5cm]{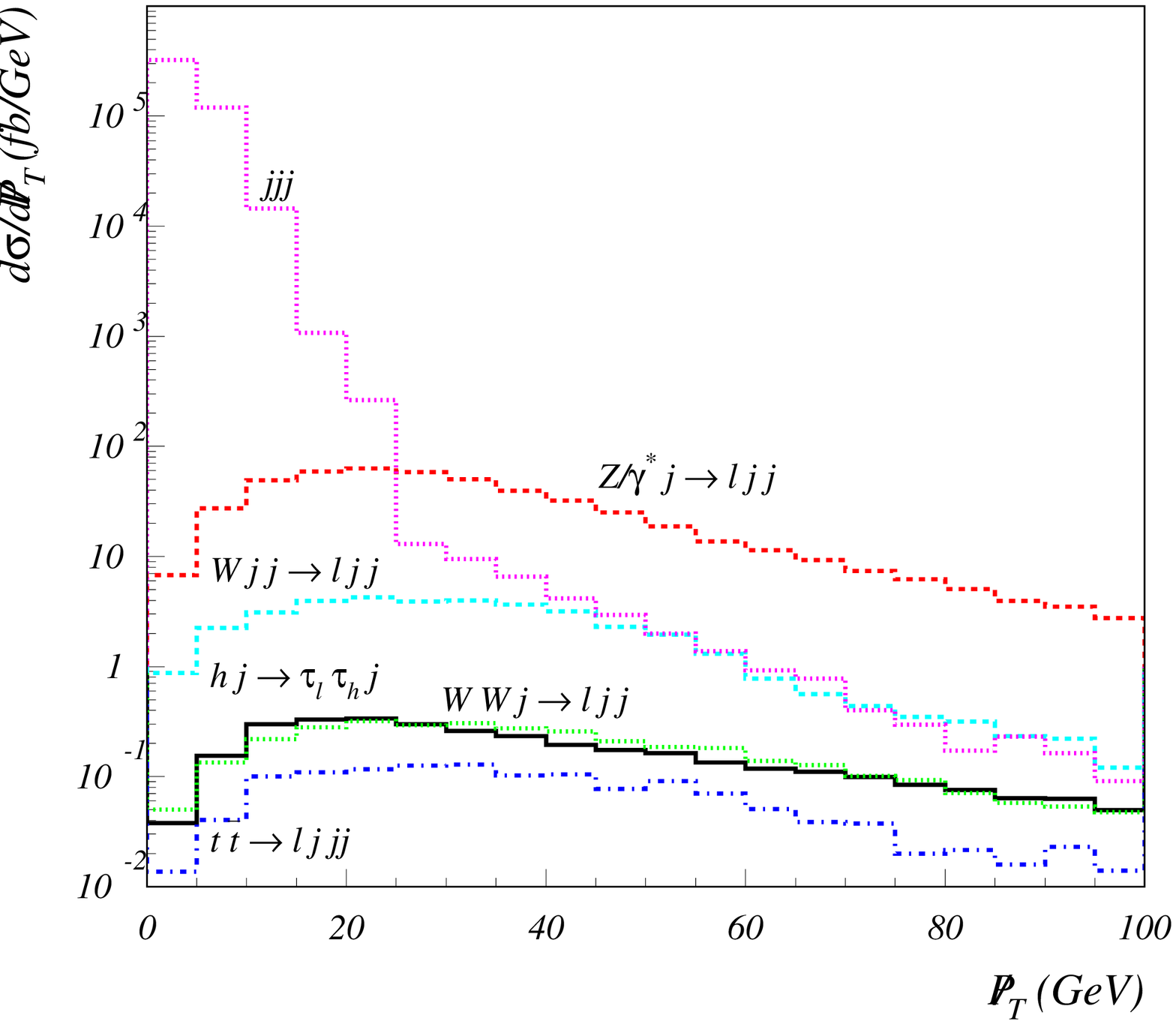}
\end{center}
\caption{Transverse missing energy distribution for signal and backgrounds for a Higgs mass of 60 $GeV$.
On the left for the $ll$ analysis and on the right for the $lj$ case. In this figure all cuts described in the text
were applied except for the missing energy cut.}
\label{fig:missing}
\end{figure}
\end{widetext}

The two other a priori very important sources of reducible background can be brought to a manageable level through a judicial cut in the transverse missing energy. This is clearly seen in fig.~\ref{fig:missing} where the transverse missing energy (imbalance of all observed momenta) distribution for a Higgs mass of 60 $GeV$ is shown. The huge QCD $pp \to jjj$ background drops five orders of magnitude as soon as we cross the 25 $GeV$ threshold for the missing energy. There is still a tail due to the leptonic decays of $c$ and $b$-quarks which involve a considerable amount of missing energy as is clear from fig.~\ref{fig:missing}. In accordance with CMS~\cite{CMS} and ATLAS studies~\cite{Aad:2009wy} we have used 0.001 as the probability of a jet faking one electron and, as explained earlier, we have taken the tau reconstruction efficiency to be 0.3 and accordingly we have used a tau rejection factor against jets as a function of the jet $p_T$~\cite{Aad:2009wy} that range from 0.01 to 0.001.

\begin{widetext}
\begin{figure}[h]
\begin{center}
\includegraphics[width=7.5cm]{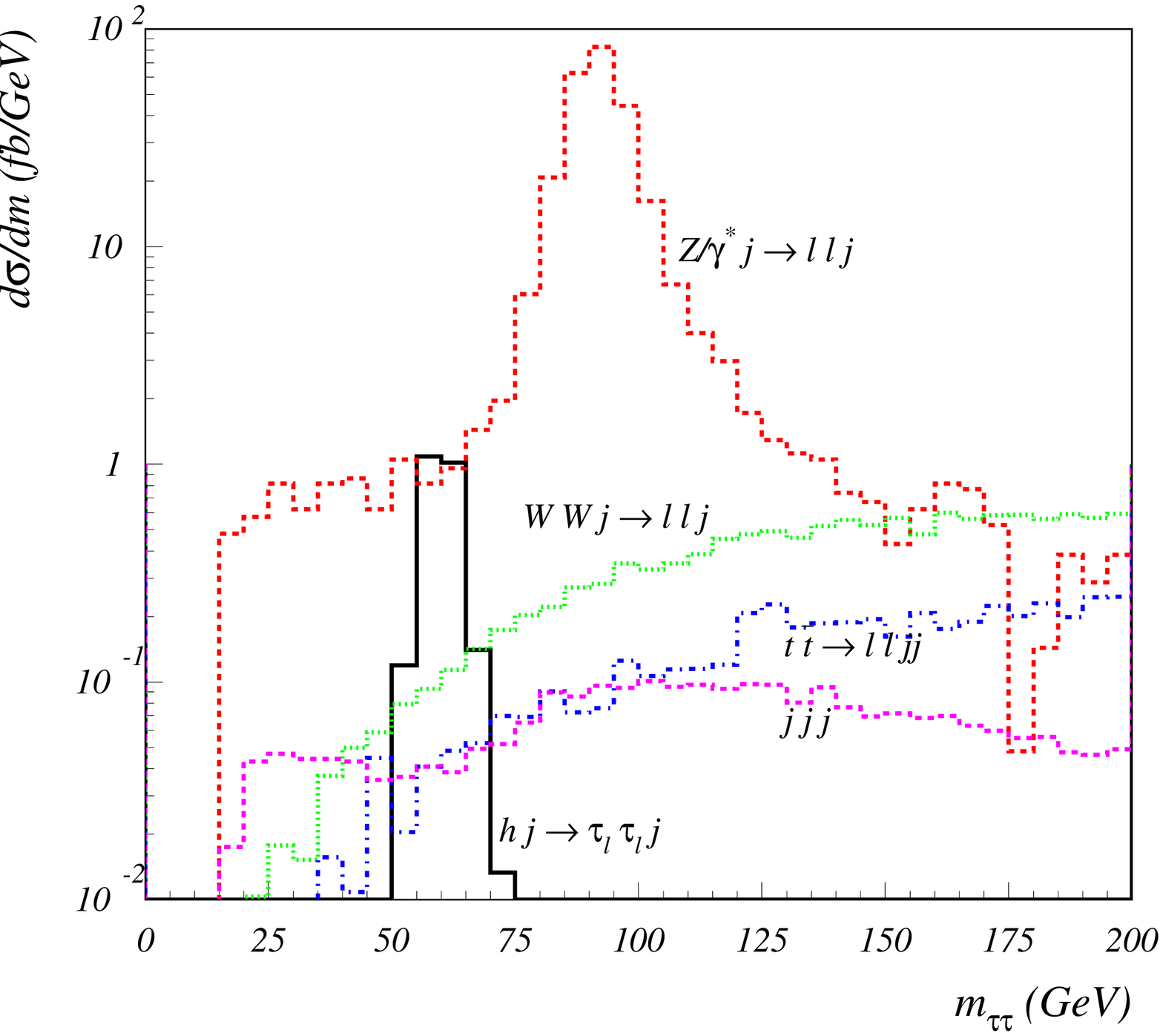}
\includegraphics[width=7.5cm]{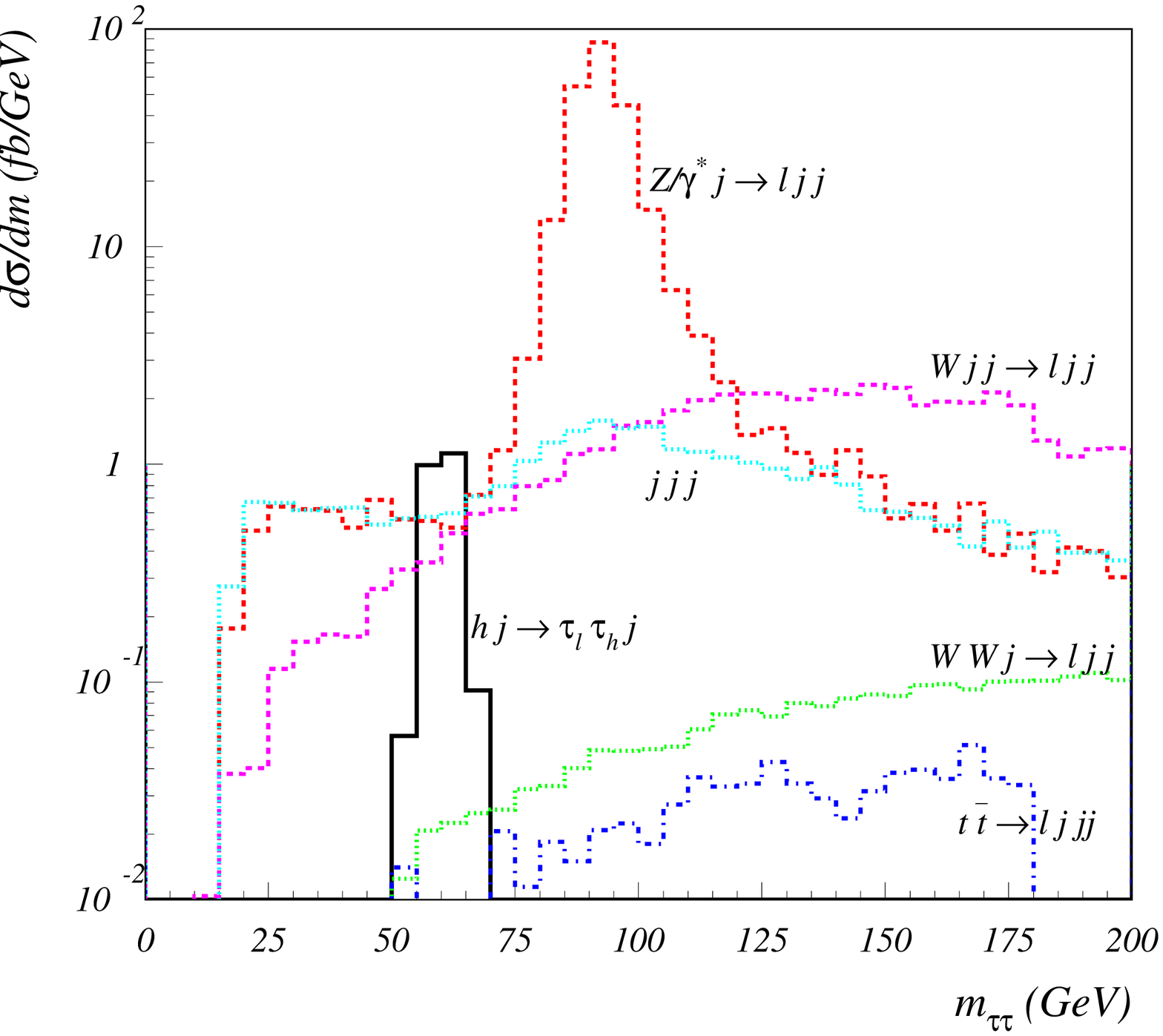}
\end{center}
\caption{Reconstructed mass $m_{\tau \tau}$ distributions for $\tau^+ \tau^-$
decaying leptonically  on the left and semi-leptonically on the right for a Higgs mass of 60 $GeV$.}

\label{fig:window}
\end{figure}
\end{widetext}

The identification of the Higgs boson signal can only be accomplished by an effective reduction of the dominating irreducible $Zj$ background. Therefore, the reconstruction of the mass peak $m_{\tau \tau}$ at $m_h$ is essential. For a more detailed discussion see~\cite{us0, Belyaev:2002zz, Ellis:1987xu}. The reconstructed mass distribution is presented in fig.~\ref{fig:window} for a Higgs mass of 60 $GeV$ after all cuts for $ll$ on the left panel and for $lj$ on the right one. In both analyses we have sharp mass peaks for the signal and also clear peaks at $m_Z$ for the $Zj$ background.

Here we summarise our analysis by specifying the following event selection procedure.
\begin{itemize}
\item We require one electron with $p_T^e > 22 \, GeV$ or one muon with $p_T^\mu > 20 \, GeV$ for
triggering purposes. An additional lepton in the event has $p^{e}_T > 15 \, GeV$ and $p^{\mu}_T
> 10 \, GeV$. A 90 \% efficiency is assumed for the reconstruction of the electron and muon and
the separation between leptons and/or jets was chosen as $\Delta R_{j(l)j(l)} > 0.4$ and $|\eta_{l} | < 3.5$ for all leptons.
\item We require that at least one jet has $p^{j}_T > \,40 GeV$ and $|\eta_{j}| < 4$.
\item We require that the hadronic tau has $p^{j}_T > \,20 GeV$ and $|\eta_{j}| < 4$.
\item We veto the event if there is an additional jet with $p^{j}_T > \,20 GeV$ and $|\eta_{j}| < 5$.
\item We apply a mass window $m_h - 15 \, GeV < m_{\tau \tau} < m_h + 15 \, GeV$.
\item Events are vetoed if the tagging jet consistent with a $b$-jet hypothesis is found with $|\eta| < 2.5$
(we assume a $b$-jet tagging efficiency of 60 \%).
\item Finally, we require the transverse missing energy to be $\slashed{E}_T > 30 \, GeV$.
\end{itemize}
%

In tab.~I
we present the SM signal and sum of all background cross sections, signal-to-background ($\sigma_{S}/\sigma_{B} $) ratios and the
significance ($\sigma_{S}/\sqrt{\sigma_{B}} $) as a function of the Higgs mass. In the first and second columns we show the results for the $ll$ analysis while columns three and four are for the $lj$ case. In the last two columns we present the combined values for the signal-to-background $\sigma_{S}/\sigma_{B} $ ratio and the significance $\sigma_{S}/\sqrt{\sigma_{B}} $. This study could still be extended to values below 20 GeV, as an experimental analysis could well be carried out for such very small values of the Higgs mass. However, the computational tools we are using here are not reliable in this Higgs mass regime, so we refrain from investigating this phenomenological possibility now.

This study could still be extended to values below 20 GeV provided the experimental analysis could be carried for very small values of the Higgs masses.

It is clear that the signal observation can be systematically challenging for the larger Higgs masses, but the values of the $\sigma_{S}/\sigma_{B} $ ratio can be improved at the expense of the significance by shrinking the Higgs mass window especially when its mass is close to the mass of the $Z$ boson. The highest significance and the highest $\sigma_{S}/\sigma_{B} $ ratio takes place for $m_h=20$ $GeV$ because it is the value farthest away from the irreducible $Z j $ background.

In tab.~\ref{tab:lumi} we present the luminosities required for a 95 \% Confidence Level (CL) exclusion, 3$\sigma$ and 5$\sigma$ discovery of a Higgs boson with SM-like Higgs couplings to the fermions at $\sqrt{s} = 14$  TeV as a function of the Higgs mass. A light Higgs boson with SM-like couplings to the fermions can be excluded at 95 \% CL in the mass range 20--60 $GeV$ with less than 1 $fb^{-1}$ of total integrated luminosity.

\begin{table}[ht]
\begin{center}
\begin{tabular}{c c c c c c c c c c c c c c c c c c c c c c c c } \hline \hline
Mass ($GeV$) &&& $\sigma_{S_{(ll)}}$ (fb)   &&&& $\sum \sigma_{B_{(ll)}}$(fb)&&&&  $\sigma_{S_{(lj)}}$ (fb)   &&&& $\sum \sigma_{B_{(lj)}}$ (fb)&&&& $\sigma_{S}/\sigma_{B}$ (\%) &&&& $\sigma_{S}/\sqrt{\sigma_{B}}$  ($\sqrt{fb}$ )  \\
\hline
20    &&&  11.6    &&&&    14.1     &&&&  5.4      &&&&    28.8    &&&& 84.1   &&&&  3.24  \\ \hline
30    &&&  11.8    &&&&    23.0     &&&&  9.9      &&&&    35.4    &&&& 58.4   &&&&  2.97  \\ \hline
40    &&&  11.5    &&&&    27.1     &&&&  10.6     &&&&    41.2    &&&& 49.7   &&&&  2.76  \\ \hline
50    &&&  11.3    &&&&    30.4     &&&&  10.9     &&&&    47.9    &&&& 43.4   &&&&  2.58  \\ \hline
60    &&&  11.9    &&&&    41.2     &&&&  11.3     &&&&    63.3    &&&& 34.0   &&&&  2.34  \\ \hline
70    &&&  13.0    &&&&    169.5    &&&&  12.0     &&&&    149.0   &&&& 11.2   &&&&  1.41  \\ \hline
80    &&&  13.8    &&&&    890.0    &&&&  12.9     &&&&    856.2   &&&& 2.2    &&&&  0.64  \\ \hline
90    &&&  14.4    &&&&    1178.3   &&&&  14.1     &&&&    1145.6  &&&& 1.7    &&&&  0.60  \\ \hline
100   &&&  14.9    &&&&    1124.7   &&&&  15.5     &&&&    1142.6  &&&& 1.9    &&&&  0.64  \\ \hline \hline
\end{tabular}
\caption{Cross sections for signal and sum of all backgrounds after all cuts as a function of the Higgs mass for a Higgs boson with SM-like couplings to the fermions. In the first and second columns we show the results for the $ll$ analysis while columns three and four are for the $lj$ case. In the last two columns we present the combined values for $\sigma_{S}/\sigma_{B}$ and $\sigma_{S}/\sqrt{\sigma_{B}}$, summed under quadrature. The analysis was done for masses between 20 and 100 $GeV$.}
\end{center}
\label{tab:sigbac}
\end{table}
\begin{table}[ht]
\begin{center}
\begin{tabular}{c c c c c c c c c c c c c c c c c c c} \hline \hline
Mass ($GeV$) &&& 95 \% CL exclusion  $L~(fb^{-1})$ &&&& 3$\sigma$ discovery $L~ (fb^{-1})$&&&&  5$\sigma$ discovery $L~ (fb^{-1})$ \\
\hline
20        &&&  0.38     &&&&    0.86  &&&&   2.38   \\ \hline
30        &&&  0.45     &&&&    1.02  &&&&   2.84   \\ \hline
40        &&&  0.53     &&&&    1.18  &&&&   3.28   \\ \hline
50        &&&  0.60     &&&&    1.35  &&&&   3.76   \\ \hline
60        &&&  0.73     &&&&    1.64  &&&&   4.56   \\ \hline
70        &&&  2.02     &&&&    4.56  &&&&   12.7   \\ \hline
80        &&&  9.76     &&&&    22.0  &&&&   61.0  \\ \hline
90        &&&  11.4     &&&&    25.7  &&&&   71.3  \\ \hline
100        &&& 9.87     &&&&    22.2  &&&&   62.0   \\ \hline \hline
\end{tabular}
\caption{Integrated luminosities needed to reach a 95 \% CL exclusion, 3$\sigma$ and 5$\sigma$ discovery for a Higgs boson with SM-like
couplings to the fermions, at the LHC. Luminosities shown are for the combined results of the two analyses
(leptonic and semi-leptonic final states) - signal to background ratio and sensitivities summed under quadrature.}
\end{center}
\label{tab:lumi}
\end{table}

\section{Extensions of the Higgs sector}
\label{sec:ext}

Which are the simplest extensions of the Higgs sector of the SM that can accommodate a very light Higgs boson? One can write an extensive list of models with very light (pseudo)scalars - it is enough to enlarge the parameter space by adding an arbitrary number of fields to accomplish such a goal. However, we want these models to reproduce the SM results and to have a high predictive power at the LHC. For simplicity we will restrict ourselves to models where Charge and Parity
(CP) is conserved in the Higgs sector and where natural flavour conservation is assumed~\cite{Glashow}. In the SM, the Higgs couplings to gauge bosons are fixed by the gauge structure and the Higgs Vacuum Expectation Value (VEV). One of the simplest extensions of the SM scalar sector is to add a neutral singlet
(i.e., $I=0$ and $Y=0$), where $I$ is the isospin and $Y$ is the hypercharge. Such a singlet does not couple to gauge bosons nor does it couple to fermions. The CP-even component of this singlet can mix with the CP-even Higgs field from the doublet. Therefore, the coupling to fermions and gauge bosons can only change due to the rotation angle related to this mixing. Calling the rotation angle $f (\chi)$, the couplings are then just redefined as
\begin{equation}
g_{VVh}^{SM} \to f (\chi) \, g_{VVh}^{SM},  \qquad \qquad  g_{ffh}^{SM} \to f (\chi) \, g_{ffh}^{SM},
\label{eq:singlet}
\end{equation}
where $V$ stands for a gauge boson (i.e., $V=W,Z$) and $f$ is a generic fermion~\footnote{If only one singlet is added, $f (\chi) = \sin \chi$ (or $\cos \chi$) depending on how one defines the rotation angle.}. In this case, having a light Higgs implies that $f (\chi)  \ll 1$ and therefore not only the searches based on production and decay processes that proceed via couplings with gauge bosons will yield negligible rates but the same is true for the ones that rely on fermion-Higgs couplings. Such a light Higgs state $h$ could only be produced then in processes involving self-Higgs couplings like for example in the gluon fusion
or vector boson fusion reactions, $gg \to H \to hh$ or in $qq \to qqH \to qqhh$,
respectively. In this scenario, the other CP-even Higgs boson, $H$, is SM-like in its couplings to vector bosons and to fermions. As for decays, the light Higgs BRs are the SM ones because all Higgs couplings fermions and gauge bosons are rescaled by the same factor, with the decay to light fermions ($b$'s, $c$'s and $\tau$'s) dominating by virtue of the small Higgs mass that the gauge boson decays are not open yet. Obviously, this scenario cannot be studied with the analysis presented in this work because of a negligible production rate for the $h$ state.

The next step is to add one doublet to the SM to obtain what is known as a 2HDM. Here we will not be concerned about any specific 2HDM potential - we just require CP conservation in the Higgs sector. The couplings to gauge bosons are universal and we define the coupling of gauge bosons to the lightest Higgs as $\sin (\beta - \alpha) \, g_{SM}$, where $\beta$ is the mixing angle in the CP-odd and charged sectors ($\tan \beta$ is also the ratio of the Higgs VEVs) and $\alpha$ is the mixing angle in the CP-even Higgs sector. The Yukawa Lagrangian can be built in four different and independent ways~\cite{catalogue} if Flavour Changing Neutral Currents (FCNCs) are to be avoided. Regarding the Yukawa Lagrangian, there are two clearly different scenarios to be considered. The first one is a SM-like scenario where only one doublet, say $\phi_2$, gives mass to all fermions usually referred to as type I model. The second class of models is the one where both doublets participate in the mass generation process. With natural flavour conservation, one can build the following models: type II is the model where $\phi_2$ couples to up-type quarks and $\phi_1$ couples to down-type quarks and leptons; in a type III model $\phi_2$ couples to up-type quarks and to leptons and $\phi_1$ couples to down-type quarks; a type IV model is instead built such that $\phi_2$ couples to all quarks and $\phi_1$ couples to all leptons. We present all Yukawa couplings in appendix A. Adding extra neutral singlets to 2HDMs amounts to a redefinition of the couplings to the SM particles equivalent to~(\ref{eq:singlet}), that is
\begin{equation}
g_{VVh}^{\rm 2HDM} \to f_{\chi_i} \, g_{VVh}^{\rm 2HDM},  \qquad \qquad  g_{ffh}^{\rm 2HDM} \to f_{\chi_i} \, g_{ffh}^{\rm 2HDM},
\label{eq:singlet2}
\end{equation}
where $f_{\chi_i}$ depends now on the number of extra singlets that are added and on the explicit form of the scalar potential.

A class of models which constitute a simple extension of the 2HDM are the ones obtained by adding an arbitrary number, $n$, (we will call these models 2HDM+nD) of doublets that \textit{do not couple} to the fermions. In what follows we will follow closely the discussion in (and notation of) \cite{Barger:2009me}, where a thorough analysis of all models discussed in this work is presented. Now we have to distinguish between type I and the other types II, III and IV. In type I models, only one doublet gives mass to the fermions - one can then build a new field which is a linear combination of all remaining $(n-1)$ doublets that do not couple to the fermions. If again we choose $\phi_2$ to give mass to the fermions and $\phi_1$ as the combined field with VEVs $v_2$ and $v_1$, respectively, we will have
\begin{equation}
v_1^2 + v_2^2 = v^2 \omega^2,   \qquad \qquad 0 < \omega \leq 1,
\end{equation}
where $\omega$ is a function of the remaining $(n-1)$ VEVs. $\omega = 1$ is the 2HDM case while $\omega <1 $ is a signal that a non-zero VEV is carried by the linear combination of the $(n-1)$ fields orthogonal to the light Higgs state.

The situation is slightly more complicated for models II, III and IV. Taking Model II as an example and following~\cite{Barger:2009me} we examine the case where just one more doublet is added, and parametrise the mixing with the extra doublet in terms of an angle $\theta$,
\begin{equation}
h= \cos \theta h' + \sin \theta h_0,
\end{equation}
where $h'$ is the usual 2HDM lightest CP-even Higgs boson defined as $h'= \cos \alpha \phi_u - \sin \alpha \phi_d$ in terms of the original doublets $\Phi_u$ and $\Phi_d$ responsible for giving mass to the fermions; $h_0$ is the CP-even state of the new doublet $\phi_0$ that does not participate in the process of mass generation. There is no mixing between the two fields when $\sin \theta =0$ and in this case $h = h'$. With the usual definition for $\tan \beta = v_2/v_1$, we define $\cos \Omega =\sqrt{(v_1^2 + v_2^2)}/v$ and $\sin \Omega = v_0 /v$, with $0 \leq \Omega < \pi/2$ to write the couplings to gauge bosons as
\begin{equation}
g_{VVh} = (\cos \Omega \cos \theta \sin (\beta - \alpha) + \sin \Omega \sin \theta) \, g_{VVh}^{SM}
\end{equation}
and to fermions
\begin{equation}
g_{\bar{u} u h} = \frac{\cos \theta}{\cos \Omega} \frac{\cos \alpha}{\sin \beta} \, g_{ffh}^{SM}, \qquad \qquad g_{\bar{d} d h} = g_{\bar{l} l h} = - \frac{\cos \theta}{\cos \Omega} \frac{\sin \alpha}{\cos \beta} \, g_{ffh}^{SM}.
\end{equation}
From the physical point of view, adding more doublets will not bring anything new to the particular case we are discussing. The same is true if we add an arbitrary number of singlets - the effect is the same as for the SM - to reduce all Higgs couplings to the 2HDM fields by the same amount.

Finally we discuss the case where the fermions masses arise from couplings to three different Higgs doublets. We restrict our study to the democratic model described and constrained in ~\cite{Grossman:1994jb}, where up-type quarks, down-type quarks and charged leptons all get their mass from a different doublet. This model is known as the democratic 3HDM and will be represented by 3HDM(D). Following~\cite{Barger:2009me} we define $\tan \beta = v_u/v_d$ and $\cos \Omega =\sqrt{(v_u^2 + v_d^2 )}/v$, $\sin \Omega = v_l /v$, where $v_u$, $v_d$ and $v_l$ are the VEVs of the doublets that couple to the up-quarks, down-quarks and charged leptons, respectively. With these definitions the couplings to gauge bosons are
\begin{equation}
g_{VVh} = (\cos \Omega \cos \theta \sin (\beta - \alpha) + \sin \Omega \sin \theta) \, g_{VVh}^{SM}
\end{equation}
while the ones to fermions can be written as
\begin{equation}
g_{\bar{u} u h} = \frac{\cos \theta}{\cos \Omega} \frac{\cos \alpha}{\sin \beta} \, g_{ffh}^{SM}, \qquad  g_{\bar{d} d h} =  - \frac{\cos \theta}{\cos \Omega} \frac{\sin \alpha}{\cos \beta} \, g_{ffh}^{SM}, \qquad  g_{\bar{l} l h} =  \frac{\sin \theta}{\sin \Omega} \, g_{ffh}^{SM}.
\label{eq:3HDMf}
\end{equation}
Again, one can add more singlets and doublets to the democratic 3HDM but this will just increase our freedom to have a light Higgs boson. We refer the reader to~\cite{Barger:2009me} for a discussion on extensions of the 3HDM.

In tab.~\ref{tab:models} we present for each model the cross sections for the processes $e^+ e^- \to Z h$ and $pp \to gh$ relative to the respective SM cross sections. In the last column, the BR$(h \to \tau^+ \tau^-)$ relative to the SM one is shown as a function of the SM BR to up-quarks, down-quarks and charged leptons. It should be noted that while the expressions for the cross sections are exact, the ones for the BRs are only truly accurate when the Higgs decay to gluons is negligible or when it proceeds mainly through a top loop (like in the SM) in which case it can easily be included in the expression using the respective $g_u$ coupling.
\begin{table}[ht]
\begin{center}
\begin{tabular}{c c c c c c c c c c c c c c c c} \hline \hline
Model &&& $\bar{\sigma} (e^+ e^- \to Z h)$   &&&& $\bar{\sigma} (gg \to gh)$ &&&& $\overline{BR} (h \to \tau^+ \tau^-)$ \\ \hline
2HDMI &&&  $\sin^2 (\beta - \alpha) $   &&&&   $\frac{\cos^2 \alpha}{\sin^2 \beta}$   &&&&   $\approx$ 1    \\ \hline
2HDMI+nD   &&&  $\omega^2 \sin^2 (\beta - \alpha) $   &&&&    $\frac{1}{\omega^2} \frac{\cos^2 \alpha}{\sin^2 \beta}$  &&&&   $\approx$ 1   \\ \hline
2HDMII       &&&  $\sin^2 (\beta - \alpha) $       &&&&    $F_{loop}$  &&&&  $[1 + (\frac{g_u^2}{g_l^2}-1)BR_u ]^{-1}$  \\ \hline
2HDMII+nD  &&&  $(\cos \Omega \cos \theta \sin (\beta - \alpha) + \sin \Omega \sin \theta)^2$       &&&&    $ \frac{\cos^2 \theta}{\cos^2 \Omega} \, F_{loop}$     &&&&   $[1 + (\frac{g_u^2}{g_l^2}-1)BR_u ]^{-1}$  \\ \hline
2HDMIII      &&&  $\sin^2 (\beta - \alpha) $      &&&&    $F_{loop}$   &&&&  $[1 + (\frac{g_d^2}{g_l^2}-1)BR_d ]^{-1}$  \\ \hline
2HDMIII+nD &&&  $(\cos \Omega \cos \theta \sin (\beta - \alpha) + \sin \Omega \sin \theta)^2$      &&&&  $\frac{\cos^2 \theta}{\cos^2 \Omega}  \, F_{loop}$    &&&&  $[1 +(\frac{g_d^2}{g_l^2}-1)BR_d ]^{-1}$ \\ \hline
2HDMIV       &&&  $\sin^2 (\beta - \alpha) $      &&&&    $\frac{\cos^2 \alpha}{\sin^2 \beta}$   &&&&   $[1 + (\frac{g_u^2}{g_l^2}-1) \, (1 - BR_{\tau})  ]^{-1}$ \\ \hline
2HDMIV+nD  &&&  $(\cos \Omega \cos \theta \sin (\beta - \alpha) + \sin \Omega \sin \theta)^2$      &&&&    $\frac{\cos^2 \theta}{\cos^2 \Omega} \frac{\cos^2 \alpha}{\sin^2 \beta}$   &&&&   $[1 + (\frac{g_u^2}{g_l^2}-1) \, (1 - BR_{\tau}) ]^{-1}$  \\ \hline
3HDM (D)      &&&  $(\cos \Omega \cos \theta \sin (\beta - \alpha) + \sin \Omega \sin \theta)^2$      &&&&    $\frac{\cos^2 \theta}{\cos^2 \Omega} \, F_{loop}$    &&&&   $[1 + (\frac{g_u^2}{g_l^2}-1)BR_u + (\frac{g_d^2}{g_l^2}-1)BR_d ]^{-1}$  \\ \hline \hline
\end{tabular}
\caption{Cross sections for $e^+ e^- \to Z h$ and $gg \to gh$ relative to the respective SM cross sections for the models discussed in the text. In the last column, the BR$(h \to \tau^+ \tau^-)$ relative to the SM one is shown; in each row, $g_i$ refers to the model on that particular row. For 2HDM the coupling are presented in the appendix and moreover, adding doublets will not alter the $g_i$ ratios therefore expressions for 2HDM and 2HDM+nD are the same. In the case of the 3HDM, the Yukawa couplings are shown in Eq. (\ref{eq:3HDMf}).}
\end{center}
\label{tab:models}
\end{table}

The function $F_{loop}$ represents the loop contribution which cannot be written as a function of the SM cross section. In the SM, the top loop contribution dominates over the bottom loop one by a factor $(m_t/m_b)^2$. This is also true in models type I and IV and their extensions. In all other models the factor that multiplies the top loop is different from the one that multiplies the bottom loop. We write this function symbolically as
\begin{equation}
F_{loop}=|\frac{\cos \alpha}{\sin \beta} \, t_{loop}^{SM}-\frac{\sin \alpha}{\cos \beta} \, b_{loop}^{SM}|^2 \, .
\end{equation}
In this case we cannot use the SM results and the process has to be recalculated.

\section{Experimental and theoretical bounds}
\label{sec:bounds}
In this section we present an overview on the bounds of the extensions of the Higgs sector discussed in the previous section. We start by noting that we would like to keep this study as general as possible. Hence, we will disregard bounds that require a knowledge about the specific Higgs potential being used. This means that only the bounds stemming from couplings to gauge bosons and to fermions will be used. We start with the lightest neutral Higgs state. As discussed in the introduction, a SM Higgs boson with a mass below 114 $GeV$ was excluded by the LEP experiments. What we are interested in this work is to know under what circumstances a light CP-even Higgs boson from a more general model could have escaped detection at LEP. All topological searches based on the processes $e^+ e^- \rightarrow H_1 Z$ and $e^+ e^- \rightarrow H_1 H_2$, where $H_1$ can be any CP-even Higgs boson and $H_2$ can be either a CP-even or a CP-odd Higgs boson were presented in~\cite{Schael:2006cr}.  Considering a scenario where all other neutral bosons are heavy enough to have eluded searches based on the various $H_1H_2$ combinations in the generic process $e^+ e^- \rightarrow H_1 H_2$, we just need to be concerned with the bounds from $e^+ e^- \rightarrow H_1 Z$. Therefore, in this study, the masses of the remaining Higgs bosons have no bearing in our analysis. For the pure 2HDMs, the smaller $\sin (\beta-\alpha)$ is the more the coupling $ZZh$ becomes negligible. Therefore, a very light Higgs could only have escaped detection in a parameter region where $\sin (\beta-\alpha) \to 0$. Depending on the Yukawa model chosen, there are two bounds we have to take into account from direct searches: one that comes from
\begin{equation}
\frac{\sigma(e^+e^-\to {H_1^{\rm 2HDM} Z)} \, {\rm BR} (H_1^{\rm 2HDM} \rightarrow b \bar{b})}{\sigma(e^+e^-\to
{H_1^{\rm SM} Z}) \, {\rm BR} (H_1^{\rm SM} \rightarrow b \bar{b})} \,  \, = \, \sin^2 (\alpha - \beta) \, \frac{{\rm BR} (H_1^{\rm
2HDM} \rightarrow b \bar{b})}{{\rm BR} (H_1^{\rm SM} \rightarrow b \bar{b})} \, \, ,
\end{equation}
and the other one originates from the process
\begin{equation}
\frac{\sigma(e^+e^-\to {H_1^{\rm 2HDM} Z)} \, {\rm BR} (H_1^{\rm
2HDM} \rightarrow \tau^+ \tau^-)}{\sigma(e^+e^-\to
{H_1^{\rm SM} Z}) \, {\rm BR} (H_1^{\rm
SM} \rightarrow \tau^+ \tau^-)}  \, = \, \sin^2 (\alpha - \beta) \, \frac{{\rm BR} (H_1^{\rm
2HDM} \rightarrow \tau^+ \tau^-)}{{\rm BR} (H_1^{\rm
SM} \rightarrow \tau^+ \tau^-)} \, \, .
\end{equation}
The reason being that, whatever the values of the 2HDM parameters chosen are, the main decays for a Higgs in the mass region below $\sim$ 100 $GeV$ are to $b \bar{b}$ and $\tau^+ \tau^-$ if decays to other Higgs are not kinematically allowed. The BRs involved depend mainly on the values of $\alpha$, $\tan \beta$ and $m_h$ that are constrained to obey
\begin{equation}
\sin^2 (\beta - \alpha) \, \frac{{\rm BR}_{th}^j (h \to b \bar{b} \, (\tau^+ \tau^-))}{{\rm BR}_{th}^{SM} (h \to b \bar{b} \, (\tau^+ \tau^-))} < \left( \frac{\sigma^{\rm 2HDM}}{\sigma^{SM}} (h \to b \bar{b}\, (\tau^+ \tau^-))\right)_{exp}
\end{equation}
where $j=$ I, II, III, IV, BR$_{th}$ is the theoretical 2HDM BR and the subscript $exp$ stands for the experimentally measured value. Note that the limit cannot be shown in the ($(\beta - \alpha)$, $m_h$) plane because of the angle dependence of the Higgs BRs to $b \bar{b}$ and to $\tau^+ \tau^-$. It is straightforward to check that when $\sin (\beta - \alpha) \approx 0.1$ there is essentially no bound on the lightest Higgs boson mass. To be more precise, in the above mass range, a $100\, \%$ BR to $b \bar{b}$ forces $\sin (\beta - \alpha) \gtrsim 0.13$ while a $100\, \%$ BR to $\tau^+ \tau^-$ implies $\sin (\beta - \alpha) \gtrsim 0.16$ regardless of the Higgs mass. Above these values the limit on $\sin (\beta - \alpha)$ strongly depends on the value of the Higgs mass. Taking as an example Model II, for $\sin (\beta - \alpha) \approx 0.2$ the limit immediately jumps to $m_h > 75.6$ $GeV$. We will therefore use $\sin (\beta - \alpha) = 0.1$ as a benchmark value. Note that in the limit $\sin (\beta - \alpha) = 0$ we would have a light gaugephobic~\cite{Pois:1993ay} Higgs boson which is again not experimentally excluded. Because $\tan \beta$ is an important parameter in our analysis we note that for a Higgs mass of 40 $GeV$  for $\tan \beta = 1$ we have $0.57 \leq \sin \alpha \leq 0.82$ while if $\tan \beta = 30$ we have $0.98 \leq \sin \alpha \leq 1$. For larger masses the bounds are obviously relaxed but the most important conclusion is that the allowed values of $\sin \alpha$ are always positive and for large $\tan \beta$ we have $\sin \alpha \approx 1$. It is important to note that these bounds do not depend on the total number of (pseudo)scalars in the model under consideration. In the reminder of this section we will discuss the limits that do depend on the number of Higgs states.

The most restrictive bound for a light scalar is the one coming from the muon anomalous magnetic moment $(g-2)_\mu$~\cite{g-2}. A very detailed account on the subject, including present status of experiment and theory, can be found in~\cite{Jegerlehner:2009ry}. A detailed study of the new physics contributions would force us to redo the calculations and subtract the diagrams where a SM Higgs takes part. In what follows we will restrict our discussion to the pure 2HDM case - all further extensions of the scalar sector will add more freedom to the model, hence they would be much more unconstrained.

\begin{widetext}
\begin{figure}[h]
\begin{center}
\includegraphics[width=7.61cm]{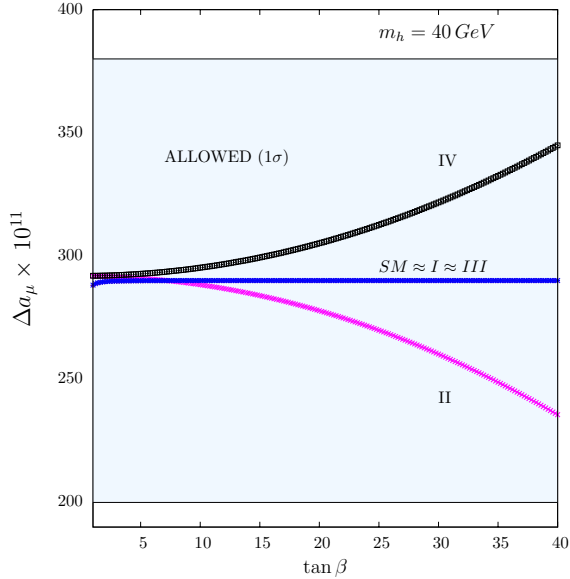}
\end{center}
\caption{$\Delta a_{\mu}= (a^{exp}_{\mu}-a^{th}_{\mu}) + a^{th-2HDM}_{\mu}$ as a function of $\tan \beta$ for $m_h=$ 40 $GeV$ and $m_A=200$ $GeV$.}
\label{fig:gm2}
\end{figure}
\end{widetext}

There are two important contributions from extended models to the muon anomalous magnetic moment: a one loop contribution, first calculated for the 2HDM in~\cite{Haber:1978jt}, and the two loop Barr-Zee contribution~\cite{Barr:1990vd}. The one loop diagram for the light Higgs is proportional to $g^2_{h \mu^+ \mu^-}$ and will therefore have the SM sign which gives a positive contribution to $(g-2)_\mu$. This could help cure the present $3 \sigma$ deviation relative to the SM. However, the two loop contribution is proportional to $g_{h \mu^+ \mu^-} \, g_{h \bar{b} b}$ and $g_{h \mu^+ \mu^-} \, g_{h \bar{t} t}$ and for SM-like couplings this amounts to a negative addition that, if large, will increase the difference between theory and experiment. In fig.~\ref{fig:gm2} we plot $\Delta a_{\mu}= (a^{exp}_{\mu}-a^{th}_{\mu}) + a^{{\rm th-2HDM}}_{\mu}$ as a function of $\tan \beta$ for a light Higgs mass of 40 $GeV$ and $m_A=200$ $GeV$. We have included both the one loop and the two loop contributions and the calculation is presented for the limit $\alpha \approx \beta$ to comply with the LEP bound on the Higgs mass. We have checked that our results for model II agree with the ones presented in~\cite{Jegerlehner:2009ry} for the same limit. For small values of $\tan \beta$ the 2HDM contributions can all be safely neglected. As $\tan \beta$ grows model II makes the discrepancy between theory and experiment to grow. The most interesting scenario is the one in model IV - the leptonic 2HDM. In this model, the new contribution moves the theoretical calculation closer to the experimental result. This is a very interesting fact, from all 2HDM this is the only one that actually helps to cure the problem. Finally, new contributions in models I and III do not vary with $\tan \beta$.

There are other bounds that constrain the pure 2HDMs which deserve a brief comment. Values of $\tan \beta$ smaller than $\approx 1$ are disallowed both by the constraints coming from $R_b$ (the $b$-jet fraction in $e^+e^-\to Z\to$ jets) \cite{LEPEWWG,SLD} and from $B_q \bar{B_q}$ mixing~\cite{Oslandk}. Limits from $B_{s} \to \mu^+ \mu^-$ are not likely to affect the models discussed here - it is sufficient to take a large charged Higgs boson mass to avoid these bounds in pure 2HDM models~\cite{Bll1}. New contributions to the $\rho$ parameter stemming from Higgs states \cite{Rhoparam} have to comply with the current limits from precision measurements \cite{pdg4}: $|\delta\rho| \la 10^{-3}$. Again for the pure 2HDMs, there are limiting cases though, related to an underlying custodial symmetry,  where the extra contributions to $\delta\rho$ vanish. Since we are in the limit of very small $\sin (\beta - \alpha)$ there are two limits to consider: one is $m_h \approx m_{H^\pm}$ and $\sin (\beta - \alpha) \approx 0$ while the other is $m_{H^\pm} = m_A^{}$. As we want a light CP-even Higgs we choose $m_{H^\pm} \approx m_A$. Note however that our results only depend on the mass of the light Higgs bosons - any change in one of the other Higgs boson masses do not affect our results.

As mentioned in the introduction, there were also searches at LEP to test the Yukawa couplings~\cite{Abbiendi:2001kp,Abdallah:2004wy} based on the channels $b \bar{b} b \bar{b}$, $b \bar{b} \tau^+ \tau^-$ and $ \tau^+ \tau^- \tau^+ \tau^-$ for Higgs masses up to 50 GeV. In a pure 2HDM we are forced to be in a region where $\alpha \approx \beta$. Hence, the light Higgs coupling to fermions is either proportional to $\tan \beta$ or to $1/\tan \beta$. The limits obtained~\cite{Abbiendi:2001kp,Abdallah:2004wy} are for
\begin{equation}
\frac{g_{h f \bar{f}}^{\rm 2HDM}}{g_{h f \bar{f}}^{\rm SM}} \sqrt{BR^{\rm 2HDM} (h \to f \bar{f})}   \, ,
\end{equation}
which in the most interesting scenarios is reduced to $\tan \beta \, \sqrt{BR^{\rm 2HDM} (h \to f \bar{f})}$ - otherwise the data gives no useful bounds on the parameters of the 2HDM. The results can be easily applied to model III and model IV because for large $\tan \beta$,  $BR(h \to b \bar{b}) \approx 100 \%$ in model III and $BR(h \to \tau^+ \tau^-) \approx 100 \%$ in model IV. In any case, even for a Higgs as light as 20 GeV, the obtained bounds are for model III $\tan \beta \lesssim 21$ and for model IV $\tan \beta \lesssim 62$.

The theoretical bounds related to tree level unitarity~\cite{unit1} and vacuum stability~\cite{vac1} (boundness from below) will not influence our results either.
Finally we note that although the pure 2HDMs play a special role here because they are protected against charge and CP breaking~\cite{Ferreira:2004yd}, the same is not true for 3HDMs or for Higgs models with even more doublets~\cite{Barroso:2006pa}.

\section{Results and discussion}

\subsection{2HDM I to IV}

\begin{widetext}
\begin{figure}[h]
\begin{center}
\includegraphics[width=8.4cm]{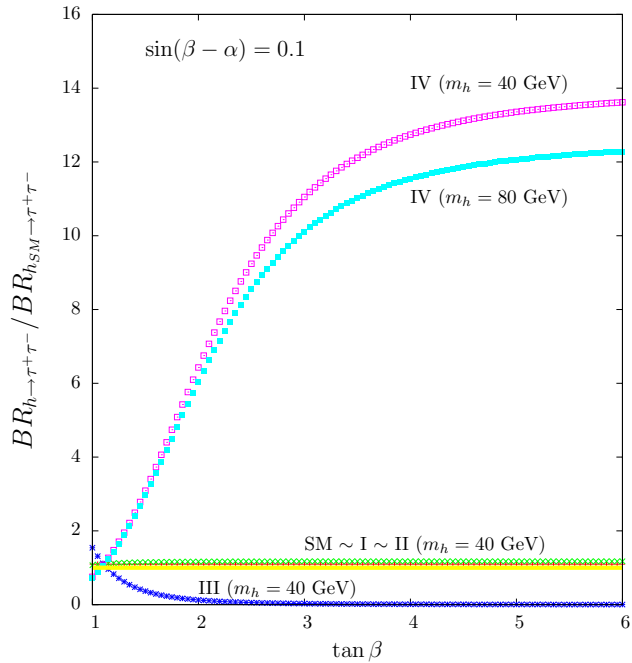}
\end{center}
\caption{Higgs BRs into $\tau^+ \tau^-$ relative to the SM one as a function of $\tan \beta$ in 2HDM Models I to IV with $\sin (\beta - \alpha)=0.1$ and $m_h=40$ $GeV$ (and also $m_h=80$ $GeV$ for model IV).}
\label{fig:BR}
\end{figure}
\end{widetext}

With the LEP bounds forcing $\sin (\beta - \alpha)$ to be small for a very light Higgs to still be allowed, the BRs and cross sections for pure 2HDMs depend almost exclusively on two parameters which we choose to be $m_h$ and $\tan \beta$. The BR$(h \to \tau^+ \tau^-)$ has a Higgs mass dependence very similar to the SM one. Therefore, big differences can only arise in the $\tan \beta$ dependence. In fig.~\ref{fig:BR} we present the Higgs BRs into $\tau^+ \tau^-$ relative to the SM one as a function of $\tan \beta$ in 2HDM Models I to IV with $\sin (\beta - \alpha)=0.1$. Model I has by construction the SM BRs except for exceptional singular points where a given parameter is null\footnote{The 2HDM fermiophobic Higgs~\cite{fermio} originates from model I by setting $\alpha= \pi/2$.}. In model II, the down quarks and charged leptons couple to the same doublet and since these are the main Higgs decays in the mass region under consideration, the behaviour is again very similar to the SM one. For the remaining two models we should distinguish the region of low $\tan \beta$ and the region of high $\tan \beta$. For $\tan \beta \approx 1$ and for model III, there will be an enhancement in BR$(h \to \tau^+ \tau^-)$ but not in the production cross section. When  $\tan \beta \gg 1$ is now model IV that sees its BR enhanced and this growth depends on the Higgs mass. Therefore the global trend is as follows: in models I and II BR$(h \to \tau^+ \tau^-)$ is SM-like in all parameter space discussed, in model III it is enhanced for small $\tan \beta$ and in model IV it grows with $\tan \beta$ when compared with the SM value (for other studies on the different Yukawa versions of the 2HDM see~\cite{oldnew}).

\begin{widetext}
\begin{figure}
\begin{center}
\includegraphics[width=7.55cm]{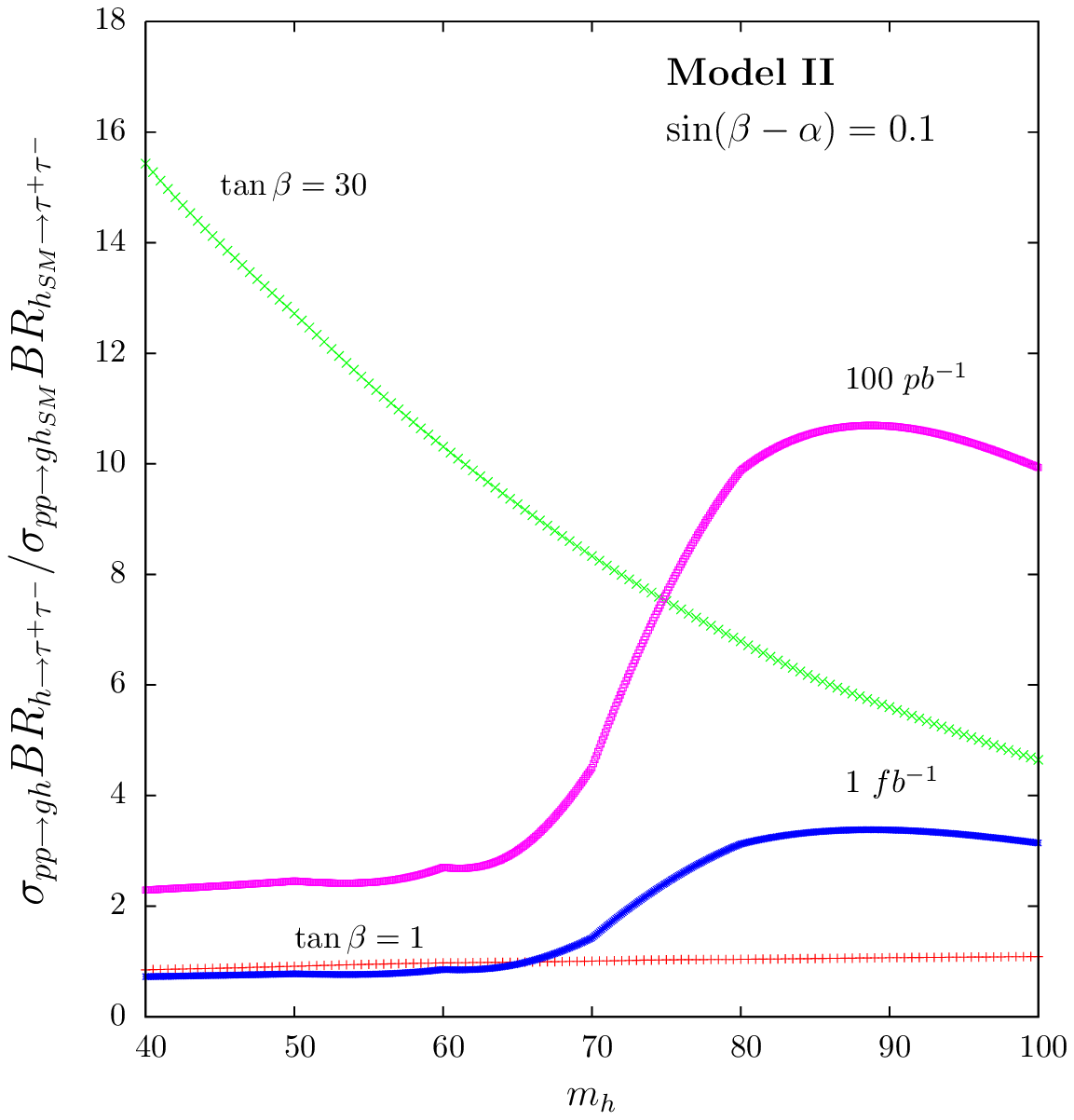}
\includegraphics[width=7.65cm]{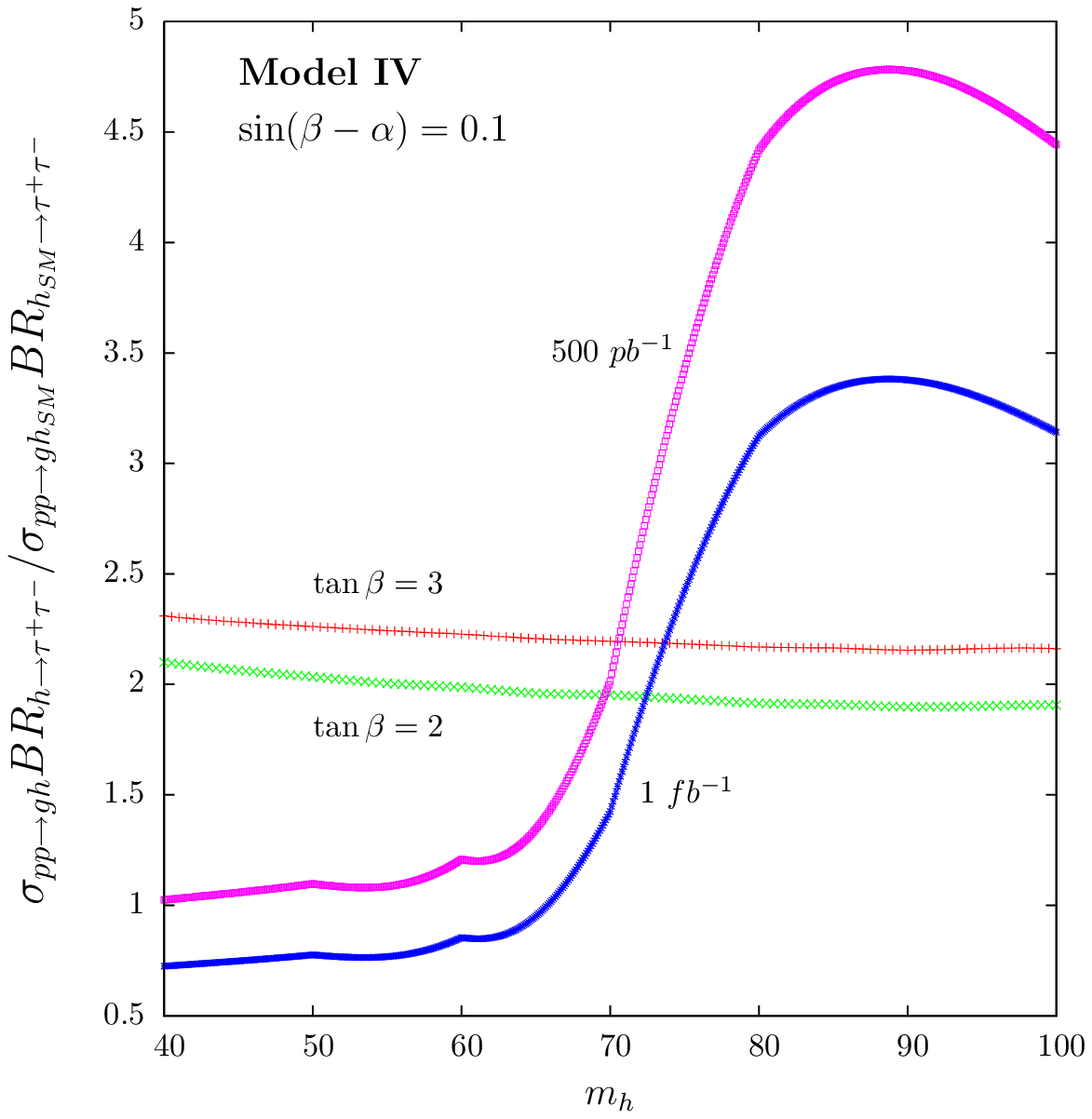}
\end{center}
\caption{In the left panel we present the ratio between $\sigma({pp \to hg}) \, {\rm{BR}}({h \to \tau^+ \tau^-})$ in model II and the SM $\sigma({pp \to h_{SM} g}) \, {\rm{BR}}({h_{SM} \to \tau^+ \tau^-})$ as a function of $m_h$ and $\tan \beta=$ 1 and 30. In the right panel we show the corresponding ratio for model IV now for $\tan \beta=$ 2 and 3. In both cases we take $\sin (\beta - \alpha)= 0.1$. We also show lines of total integrated luminosity $100 \, pb^{-1}$ and $1 \, fb^{-1}$ for model II and $500 \, pb^{-1}$ and $1 \, fb^{-1}$ for model IV.}
\label{fig:2HDM}
\end{figure}
\end{widetext}

The process $pp \to gg (qg) \to h j$ proceeds via a quark loop. As the Yukawa couplings are proportional to the quark masses, only top and bottom loops give non negligible contributions to the cross section. In the SM, the top loop is always the dominant. In our study there are two cases to consider. In models type I and IV the light Higgs couples to up and down quarks with the same strength. In this case we can write
\begin{equation}
\sigma^{\rm 2HDM} (pp \to gg \to h) = \frac{\cos^2 \alpha}{\sin^2
\beta} \quad \sigma^{SM} (pp \to gg \to h)
\end{equation}
which in turn means that, in the limit $\sin (\beta - \alpha) \to 0$, the relation is approximately
\begin{equation}
\sigma^{\rm 2HDM} (pp \to gg \to h) = \frac{1}{\tan^2 \beta}
\quad \sigma^{SM} (pp \to gg \to h).
\end{equation}
As we saw earlier, the available constraints force $\tan \beta$ to be above 1. Therefore, in models I and IV, the 2HDM cross section is the SM cross section for $\tan \beta = 1$ and then drops like $\tan^2 \beta$ with growing $\tan \beta$. This means that for $\tan \beta =10$ the cross section is 100 times smaller and even if the decay to $\tau^+ \tau^-$ reaches 100 $\%$ it will still be 10 times smaller than the corresponding SM cross section. In models type II and III the light Higgs couples to the up quarks as $\cos \alpha /\sin \beta$ and to the down quarks as $- \sin \alpha /\cos \beta$. Hence the contribution from each loop depends heavily on the value of $\tan \beta$. The larger $\tan \beta$ is the more the decay to $\tau^+ \tau^-$ becomes negligible in model III. In model II the width $\Gamma (h \to \tau^+ \tau^-)$ preserves the SM proportionality to $\Gamma (h \to b \bar{b})$ and as the cross section grows for large $\tan \beta$, the ratio $\sigma({pp \to hg}) \, {\rm{BR}}({h \to \tau^+ \tau^-})$ to the SM $\sigma({pp \to h_{SM} g}) \, {\rm{BR}}({h_{SM} \to \tau^+ \tau^-})$ will also increase. In fig.~\ref{fig:2HDM} we present this ratio for model II with $\tan \beta=1,\, 30$ (left) and for model IV with $\tan \beta =2, \, 3$ (right), as a function of the Higgs mass and with $\sin (\beta - \alpha)= 0.1$. All cross sections in this section were calculated at leading order. As stated before, we have chosen the benchmark $\sin (\beta - \alpha)=0.1 $, which is representative of a Higgs whose mass is not bounded by the available experimental data. We present luminosity lines of $1 \, fb^{-1}$ and $100 \, pb^{-1}$ for Model II and $1 \, fb^{-1}$ and $500 \, pb^{-1}$ for Model IV. These lines represent the integrated luminosity needed to exclude the model at 95 \% CL. Obviously, as the Higgs mass approaches the $Z$ boson mass, the required luminosity grows. However, there are regions of the parameter space that can be probed with less than $100 \, pb^{-1}$ of integrated luminosity with the LHC working at an energy of 14 TeV. The regions easily probed are always for the low mass region (below 60 GeV) and large $\tan \beta$ values in Model II and small to moderate values of $\tan \beta$ for Model IV.

\vspace*{0.25cm}
\subsection{Beyond 2HDMs}

\begin{widetext}
\begin{figure}
\begin{center}
\includegraphics[width=7.7cm]{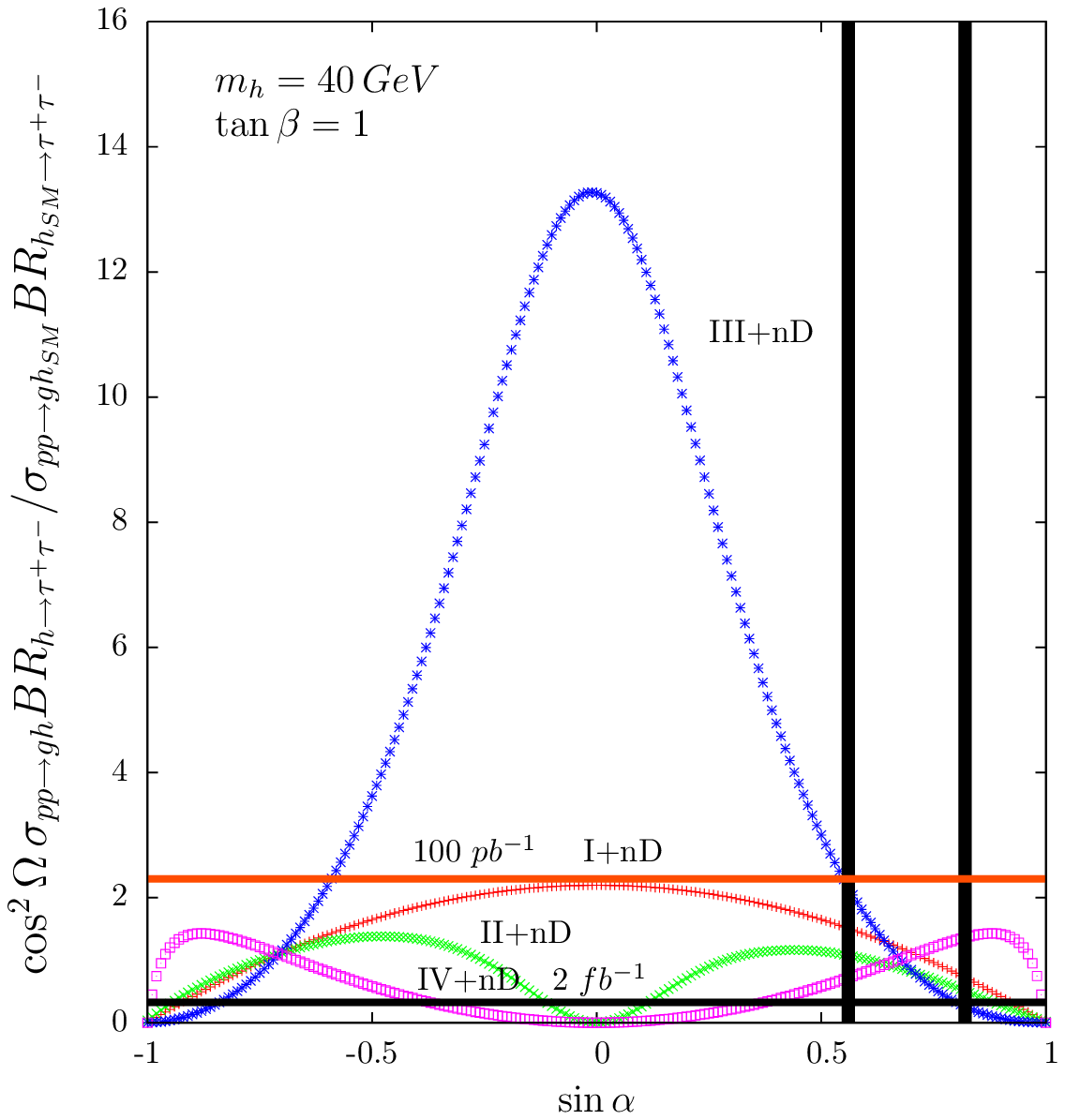}
\includegraphics[width=7.7cm]{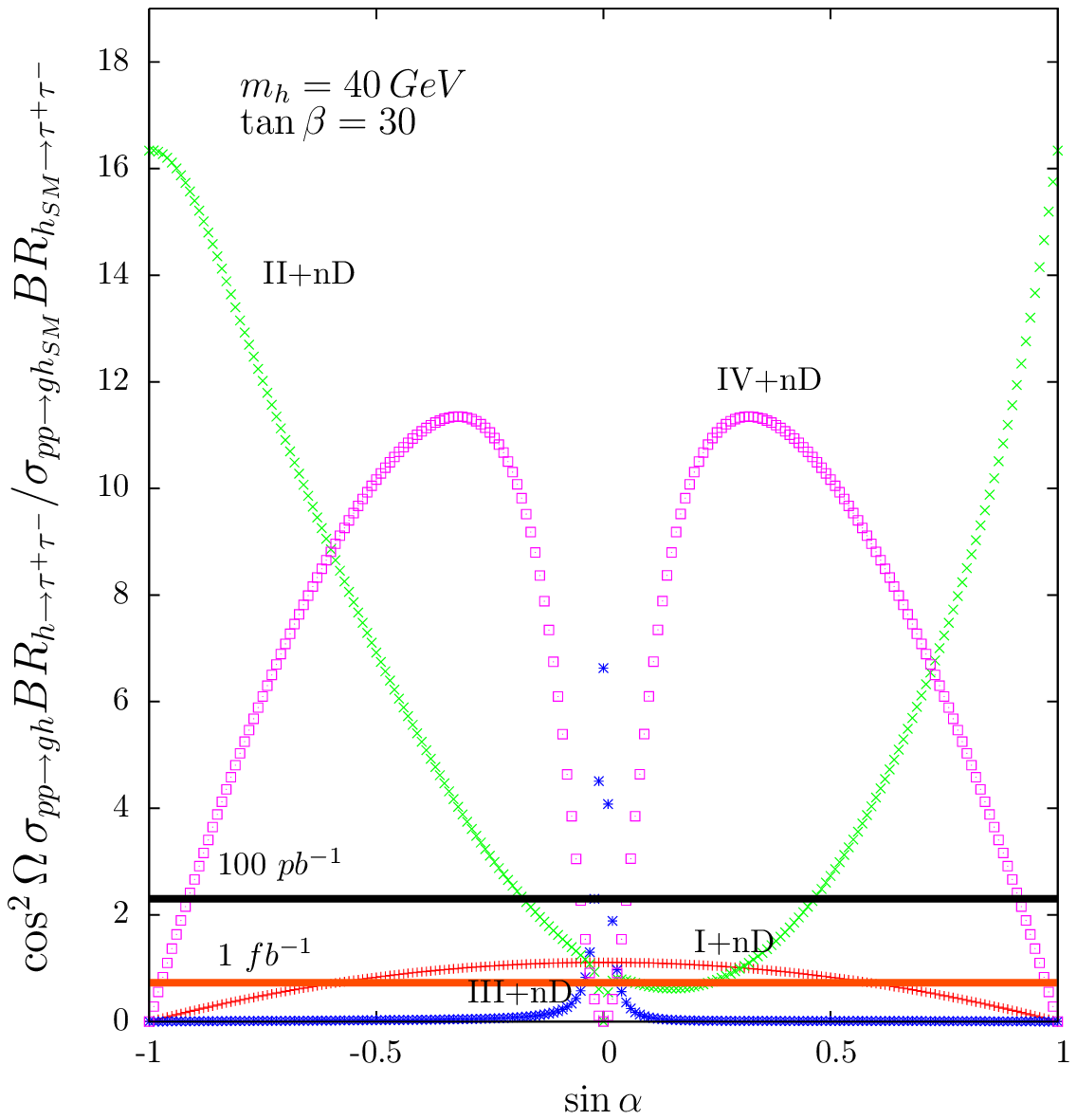}
\end{center}
\caption{In both plots the ratio between $\sigma({pp \to hg}) \, {\rm{BR}}({h \to \tau^+ \tau^-})$ and the SM $\sigma({pp \to h_{SM} g}) \, {\rm{BR}}({h_{SM} \to \tau^+ \tau^-})$ is shown, multiplied by the factor $cos^2 \Omega$ for $m_h = 40 \, GeV$ and two values of $\tan \beta$, $1$ (left) and $30$ (right) for all extensions of the 2HDM in the limit described in the text. We also present the total integrated luminosities $100 \, pb^{-1}$ and $2 \, fb^{-1}$ (left) and $100 \, pb^{-1}$ and $1 \, fb^{-1}$ (right).}
\label{fig:2HDMext}
\end{figure}
\end{widetext}

When we proceed to more general models, the first step is again to make sure the LEP bounds are respected. In tab.~\ref{tab:models} we present the Higgs couplings to gauge bosons for models with more than two doublets which is written as
\begin{equation}
(\cos \Omega \cos \theta \sin (\beta - \alpha) + \sin \Omega \sin \theta)^2
\end{equation}
where there are two new parameters to consider: $\sin \Omega$ which is a measure of the contribution of the new VEV(s) (it can come from the third doublet or from a combination of more than one) and $\sin \theta$ which determines the amount of mixing of the new CP-even field(s) to the lightest CP-even state from the 2HDM. Note that, except for the 3HDM case, just the first two doublets give mass to the fermions. There are several ways to make this quantity small enough to avoid the LEP bound. We start by considering the scenario where almost no mixing occurs between the new doublet and the remaining ones and the new VEV is maximal, which can be translated into $\cos \Omega \ll 1$ and $\sin \theta \ll 1$. In this scenario, the fermion Yukawa couplings have to be increased to give the fermions the required masses. All cross sections, independently of the Yukawa model under study, are now rescaled according to
\begin{equation}
\sigma^{\rm 2HDM} \to \frac{1}{\cos^2\Omega } \, \sigma^{\rm 2HDM}
\end{equation}
and this means that there is an enhancement already at the level of the cross section. Both $\alpha$ and $\tan \beta$ are now free to vary in all the allowed range,  provided theoretical and experimental constraints are fulfilled. In fig.~\ref{fig:2HDMext} we present the ratio between $\sigma({pp \to hg}) \, {\rm{BR}}({h \to \tau^+ \tau^-})$ and the SM $\sigma({pp \to h_{SM} g}) \, {\rm{BR}}({h_{SM} \to \tau^+ \tau^-})$, multiplied by the factor $\cos^2 \Omega$ for $m_h = 40 \, GeV$ and two values of $\tan \beta$, $1$ (left) and $30$ (right) for all Yukawa extensions of the 2HDM+nD. Note that to get the actual value of the cross section one needs to multiply it by $1/\cos^2 \Omega$ and therefore the numbers will always be larger than the ones shown in the figures. We also present several values of the total integrated luminosity needed to probe the models at 95 \% CL. In the left panel we can see that $100 \, pb^{-1}$ are enough to constraint a big portion of model III+nD while with $2 \, fb^{-1}$ just marginal regions of the models are left untested. Between the two vertical lines, the allowed region of the parameter space for the pure 2HDM is shown - where $\alpha$ is constrained due to our choice of $\sin (\beta - \alpha) = 0.1$ and $\tan \beta = 1$. In the right panel, $\tan \beta = 30$, it is now models II and IV that have the largest cross sections. We show the 95 \% CL lines for $100 \, pb^{-1}$ and $1 \, fb^{-1}$ total integrated luminosity. For large $\tan \beta$ only model III+nD and regions close to $\sin \alpha = 0$ in the other models will not be excluded with a few $fb^{-1}$ of integrated luminosity. The decay to tau pairs is negligible in this limit for the 3HDM(D).

\begin{widetext}
\begin{figure}
\begin{center}
\includegraphics[width=8cm]{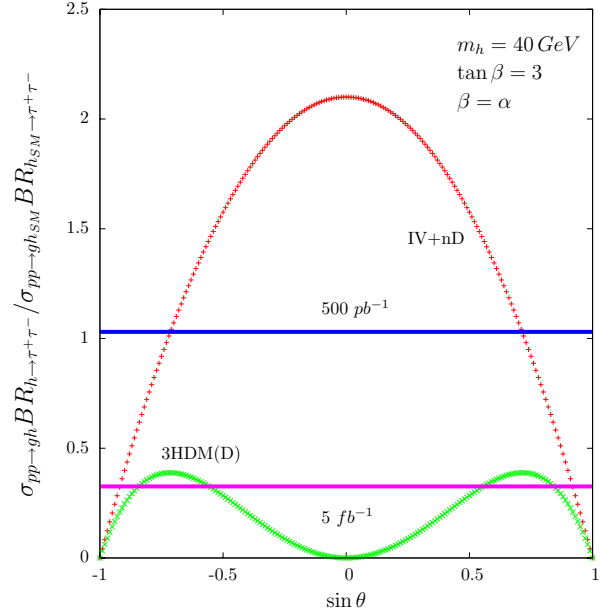}
\end{center}
\caption{Ratio between $\sigma({pp \to hg}) \, {\rm{BR}}({h \to \tau^+ \tau^-})$ and the SM $\sigma({pp \to h_{SM} g}) \, {\rm{BR}}({h_{SM} \to \tau^+ \tau^-})$ as a function of $\sin \theta$ and $\tan \beta= 3$ and $m_h = 40 \, GeV$. Models IV+nD and 3HDM(D) are compared.}
\label{fig:3HDM}
\end{figure}
\end{widetext}

We now consider the scenario $\sin \Omega \ll 1$ and $\alpha \approx \beta$ and we will explore, as an example, model IV+nD and the democratic 3HDM. Production cross sections are now rescaled as
\begin{equation}
\sigma^{\rm 2HDM} \to \cos^2 \theta  \, \sigma^{\rm 2HDM}
\end{equation}
which means that they will now be smaller than the corresponding 2HDM cross sections. In fig.~\ref{fig:3HDM} we show the ratio between $\sigma({pp \to hg}) \, {\rm{BR}}({h \to \tau^+ \tau^-})$ and the SM $\sigma({pp \to h_{SM} g}) \, {\rm{BR}}({h_{SM} \to \tau^+ \tau^-})$ as a function of $\sin \theta$ for $\tan \beta= 3$ and $m_h = 40 \, GeV$. For definiteness we take $\sin \Omega = 0.1$ and $\alpha = \beta$. The prospect of excluding a light Higgs in model IV+nD is good but even with $5 fb^{-1}$ of integrated luminosity only a small portion of the 3HDM(D) will be probed at 95 \% CL. Hence, only several years of integrated luminosity will allow us to exclude a light Higgs from a 3HDM(D) in this limit.

There are other scenarios where the LEP bound could be avoided. One is $\sin \theta << 1$ and $\alpha \approx \beta$. The first relation eliminates the 3HDM(D) light Higgs  decay to leptons. For all 2HDM+nD the cross sections are again rescaled via the factor $1/\cos^2 \Omega$ as compared to the pure 2HDM. Therefore, the results presented in fig.~\ref{fig:2HDM} will be rescaled with $1/\cos^2 \Omega$ to apply for the corresponding 2HDM+nD models. There is also the possibility of having $\sin \Omega << 1$ and $\cos \theta << 1$. Due to the second relation all production cross sections will be severely reduced. The only model that has a small chance to be probed in such a limit is the 3HDM(D) because the BR to leptons can be of the order 100 \%.

Finally, there is a completely different class of options where none of the previous limits is required. By taking
\begin{equation}
g_{VVh} = (\cos \Omega \cos \theta \sin (\beta - \alpha) + \sin \Omega \sin \theta) \, g_{VVh}^{SM} = 0
\end{equation}
and therefore
\begin{equation}
\sin (\beta - \alpha) = - \tan \Omega \tan \theta \, ,
\end{equation}
we do not need to require any special limit to avoid the LEP bound. As an example, if $\sin (\beta - \alpha) = - \tan \Omega = - \tan \theta = - 1$ the LEP bound is avoided and the lightest Higgs from 3HDM (D) will have SM-like couplings to fermions. Therefore, the SM results shown in table II, can be applied to the 3HDM (D). Note that there is no such limit in pure 2HDM cases where a small $\sin (\beta - \alpha)$ is required.


\section{Conclusions}
\label{sec:conclusions}

We have discussed the possibility of finding a very light Higgs boson at the LHC. The existence of such a very light CP-even scalar is severely constrained by the LEP bounds which assume the SM coupling for the vertex $ZZh$. If however this coupling is smaller than the SM one, the bound is relaxed and in some scenarios the bound on the Higgs mass even ceases to exist. This can be accomplished by the introduction of extra Higgs fields. We have seen that the introduction of a neutral singlet is enough to avoid the bounds but for this particular process, the production cross section becomes too small. Next, we have introduced an arbitrary number of doublets and singlets with the following restrictions: no FCNC are generated at tree level and CP is conserved. We have shown that in pure 2HDMs the limit of very small $(\beta - \alpha)$ is the only way to avoid the LEP bounds - if one further extends the scalar sector, several combinations of different limits in the parameter space lead to the same result. We have also shown that, whatever the model is, a few $fb^{-1}$ of integrated luminosity are enough to probe large portions of the associated parameter space.

With the LHC running and with the search for the Higgs boson on the way, we should ask ourselves what to do if we do not find a SM Higgs boson. It seems clear that we should turn our attention to more general potentials and in particular to the ones where a light Higgs boson is allowed. However, even if a Higgs boson is found and even if looks very much like the SM Higgs boson, we should make sure that we did not miss any other (pseudo)scalar particle potentially present in the data. We believe this work is a very important contribution to achieve such a goal.

\appendix
\section{Yukawa couplings of a 2HDM with flavour conservation}
\noindent
In this Appendix we present the Feynman rules for the 2HDM Yukawa
couplings. Hereafter, the label $u$ refers to up-type quarks and neutrinos whilst $d$ to
down-type quarks and leptons. Also notice that the Goldstone bosons couple just like in
the SM, so we do not report their fermionic interactions here. Finally,
we define $\gamma_L = (1- \gamma_5)/2$ and
$\gamma_R=(1+\gamma_5)/2$. Using notation already introduced (apart from $V_{ij}$ being
the Cabibbo-Kobayashi-Maskawa matrix element in the quark sector and equating to 1 in the
lepton case), one has (see tab. IV) the following
Feynman rules:

\begin{table}[!t]
\begin{center}
\begin{tabular}{c c c c c c c c } \hline \hline
           Type      &  \textbf{I} && \textbf{II} && \textbf{III} && \textbf{IV} \\ \hline
                 \hline
$\alpha_{eh}$ & $-\frac{\cos \alpha}{\sin \beta}$ && $\frac{\sin
\alpha}{\cos \beta}$ && $-\frac{\cos \alpha}{\sin \beta}$ &&
$\frac{\sin \alpha}{\cos \beta}$ \\ \hline
$\alpha_{dh}$ & $-\frac{\cos \alpha}{\sin \beta}$ && $\frac{\sin
\alpha}{\cos \beta}$ && $\frac{\sin \alpha}{\cos \beta}$ &&
$-\frac{\cos \alpha}{\sin \beta}$
 \\ \hline
$\alpha_{eH}$ & $-\frac{\sin \alpha}{\sin \beta}$ && $-\frac{\cos
\alpha}{\cos \beta}$ && $-\frac{\sin \alpha}{\sin \beta}$ &&
$-\frac{\cos \alpha}{\cos \beta}$ \\ \hline
$\alpha_{dH}$ & $-\frac{\sin \alpha}{\sin \beta}$ && $-\frac{\cos
\alpha}{\cos \beta}$ && $-\frac{\cos
\alpha}{\cos \beta}$ &&  $-\frac{\sin \alpha}{\sin \beta}$ \\
\hline
$\beta_{e}$ & $-\cot \beta$ && $\tan \beta$ && $-\cot \beta$ &&
$\tan \beta$ \\ \hline
$\beta_{d}$ & $-\cot \beta$ && $\tan \beta$ && $\tan \beta$ && $-
\cot \beta$ \\ \hline \hline
\end{tabular}\label{tab:yukawas}
\caption{Coupling constants for the fermion-scalar interactions.}
\end{center}
\end{table}

\begin{tabular}{lclclll}
$\overline{e}_i e_i h$ & & $\frac{ig}{2 M_W}\alpha_{eh} m_{e_i}$ &
$\qquad \qquad \qquad \overline{u}_i u_i G_0$ & & $-\frac{g}{2 M_W} m_{u_i} \gamma_5$ \\ \\
$\overline{u}_i u_i h$ & & $-\frac{ig}{2 M_W}\frac{\cos
\alpha}{\sin \beta} m_{u_i}$ &
$\qquad \qquad \qquad\overline{d}_i d_i G_0$ & & $\frac{g}{2 M_W} m_{d_i} \gamma_5$ \\ \\
$\overline{d}_i d_i h$ & & $\frac{ig}{2 M_W}\alpha_{dh} m_{d_i}$ &
$\qquad \qquad \qquad\overline{e}_i \nu_i H^+$ & & $\frac{ig}{2 \sqrt{2}M_W}\beta_{e} m_{e_i} (1 + \gamma_5)$ \\ \\
$\overline{e}_i e_i H$ & & $\frac{ig}{2 M_W}\alpha_{eH} m_{e_i}$ &
$\qquad \qquad \qquad\overline{u}_i d_j H^+$ & & $\frac{ig}{2 \sqrt{2}M_W} V_{ij} \left[ \beta_{d} m_{d_j} (1 + \gamma_5) + \cot \beta m_{u_i} (1 - \gamma_5) \right]$ \\ \\
$\overline{u}_i u_i H$ & & $-\frac{ig}{2 M_W}\frac{\sin
\alpha}{\sin \beta} m_{u_i}$ &
$\qquad \qquad \qquad\overline{\nu}_i e_i H^-$ & & $\frac{ig}{2 \sqrt{2}M_W}\beta_{e} m_{e_i} (1 - \gamma_5)$ \\ \\
$\overline{d}_i d_i H$ & & $\frac{ig}{2 M_W}\alpha_{dH} m_{d_i}$ &
$\qquad \qquad \qquad\overline{d}_i u_j H^-$ & & $\frac{ig}{2 \sqrt{2}M_W} V_{ij}^* \left[ \beta_{d} m_{d_i} (1 - \gamma_5) + \cot \beta m_{u_j} (1 + \gamma_5) \right]$ \\ \\
$\overline{e}_i e_i A$ & & $-\frac{g}{2 M_W}\beta_{e} m_{e_i}
\gamma_5$ &
$\qquad \qquad \qquad\overline{e}_i \nu_i G^+$ & & $-\frac{ig}{2 \sqrt{2}M_W} m_{e_i} (1 + \gamma_5)$ \\ \\
$\overline{u}_i u_i A$ & & $-\frac{g}{2 M_W} \cot \beta m_{u_i}
\gamma_5$ &
$\qquad \qquad \qquad\overline{u}_i d_j G^+$ & & $\frac{ig}{2 \sqrt{2}M_W} V_{ij} \left[ -m_{d_j} (1 + \gamma_5) + m_{u_i} (1 - \gamma_5) \right]$ \\ \\
$\overline{d}_i d_i A$ & & $-\frac{g}{2 M_W}\beta_{d} m_{d_i}
\gamma_5$ &
$\qquad \qquad \qquad\overline{\nu}_i e_i G^-$ & & $-\frac{ig}{2 \sqrt{2}M_W} m_{e_i} (1 - \gamma_5)$ \\ \\
$\overline{e}_i e_i G_0$ & & $\frac{g}{2 M_W} m_{e_i} \gamma_5$&
$\qquad \qquad \qquad\overline{d}_i u_j G^-$ & & $\frac{ig}{2 \sqrt{2}M_W}V_{ij}^* \left[ -m_{d_i} (1 - \gamma_5) + m_{u_j} (1 + \gamma_5) \right]$ \\ \\
\end{tabular}

\noindent {\bf Acknowledgments} RBG is supported by a Funda\c{c}\~ao para a Ci\^encia e Tecnologia Grant SFRH/BPD/47348/2008.
SM is financially supported in part by the scheme `Visiting Professor - Azione D - Atto Integrativo tra la
Regione Piemonte e gli Atenei Piemontesi. RS is supported by the FP7 via a Marie Curie Intra European Fellowship,
contract number PIEF-GA-2008-221707.

\end{document}